\documentclass[letterpaper]{article} 
\usepackage{aaai24}  
\usepackage{times}  
\usepackage{helvet}  
\usepackage{courier}  
\usepackage[hyphens]{url}  
\usepackage{graphicx} 
\usepackage{natbib}  
\usepackage{caption} 
\usepackage{algorithm}
\usepackage{algorithmic}

\usepackage{newfloat}
\usepackage{listings}
\DeclareCaptionStyle{ruled}{labelfont=normalfont,labelsep=colon,strut=off} 
\floatstyle{ruled}
\newfloat{listing}{tb}{lst}{}
\floatname{listing}{Listing}
\nocopyright 

\usepackage{color}
\usepackage{xcolor}
\usepackage{subcaption}
\usepackage{pifont}
\usepackage{tikz}
\usetikzlibrary{arrows.meta,chains,positioning,shapes.geometric}
\usepackage{marvosym}
\usepackage{fontawesome}
\usepackage[skins]{tcolorbox}
\DeclareUnicodeCharacter{0308}{o}

\newtcolorbox{principle}[2][]{
  enhanced,colback=white,colframe=black,coltitle=white,colbacktitle=black,
  top=7pt,bottom=0pt,left=2pt,right=2pt,
  rounded corners,  arc=2pt,
  boxrule=0.4pt,
  fonttitle=\itshape,
  attach boxed title to top left={yshift=-0.4\baselineskip-0.6pt,xshift=2mm},
  boxed title style={left=1pt,right=5pt,top=-2pt,bottom=-3pt,
    colback=black,  rounded corners, arc=1pt, before upper=\strut},
  title=#2,#1
}

\newcommand*\circled[1]{\tikz[baseline=(char.base)]{\footnotesize
    \node[shape=circle,draw,fill=black,text=white,inner sep=1pt] (char) {#1};}}

\newcommand*\sepline{%
  \begin{center}
    \rule[1ex]{.17\textwidth}{.5pt}~{\LARGE\ding{104}}~\rule[1ex]{.17\textwidth}{.5pt}
  \end{center}}

\newtheorem{definition}{Definition}[section]

\title{Pay Attention: a Call to Regulate the Attention Market and Prevent Algorithmic Emotional Governance}
\author {
    Franck Michel\textsuperscript{\rm 1},
    Fabien Gandon\textsuperscript{\rm 2}
}
\affiliations {
    \textsuperscript{\rm 1}University Côte d'Azur, CNRS, Inria\\
    \textsuperscript{\rm 2}University Côte d'Azur, Inria, CNRS\\
    franck.michel@inria.fr, fabien.gandon@inria.fr
}

\begin{document}

\maketitle

\begin{abstract}
Over the last 70 years, we, humans, have created an economic market where attention is being captured and turned into money thanks to advertising. During the last two decades, leveraging research in psychology, sociology, neuroscience and other domains, Web platforms have brought the process of capturing attention to an unprecedented scale.
With the initial commonplace goal of making targeted advertising more effective, the generalization of attention-capturing techniques and their use of cognitive biases and emotions have multiple detrimental side effects such as polarizing opinions, spreading false information and threatening public health, economies and democracies. This is clearly a case where the Web is not used for the common good and where, in fact, all its users become a vulnerable population.
This paper brings together contributions from a wide range of disciplines to analyze current practices and consequences thereof. Through a set of propositions and principles that could be used do drive further works, it calls for actions against these practices competing to capture our attention on the Web, as it would be unsustainable for a civilization to allow attention to be wasted with impunity on a world-wide scale.
\end{abstract}

\section{An Unsustainable Attention Market}
Since the advent of mass market in the 50's, media and advertisement providers have relentlessly tried to figure out effective methods to capture our attention and turn it into revenue.
During the last two decades, supported by advances in artificial intelligence (AI), major online social media and Web platforms have brought this process of capturing attention to an unprecedented scale.
Based almost exclusively on advertising revenues, their business model consists in providing free services that, in return, collect behavioral traces. This data is then used to maximize the impact of advertisements on users by \circled{1}~ensuring their mental availability at the time of being shown the advertisement, and 
\circled{2}~ensuring that the message meets their interests, beliefs and moods (i.e. targeted advertising). 
Based on research in psychology, sociology and neuroscience, several actors including online social media, games and Web platforms have engineered techniques capable of very effectively plundering our ``available brain time''~\cite{lewis_our_2017,harris_how_2019}.
We can distinguish two broad categories of such techniques.

Firstly, some techniques are explicitly designed to \textbf{leverage cognitive biases} as a means to capture attention. 
For instance, the \textit{likes} collected after posting content activate the brain's dopaminergic pathways (involved in the reward system) and tap into our need for social approval, giving ``bright dings of pseudo-pleasure''~\cite{lewis_our_2017};
notifications of smartphone applications feed our appetite for novelty and surprise such that it is difficult to resist the urge to check them;
the pull-to-refresh mechanism~\cite{lewis_our_2017}, alike slot machines, exploits the variable reward pattern whereby each time we pull down the screen we may get an update or nothing at all;
infinite scrolling (of news, posts or videos...) traps us because of our fear of missing out important information (FOMO) to the point that we can hardly break the flow;
automatic video chaining replaces a deliberate action to continue watching with a required action to stop watching, and entails a frustrating feeling of incompleteness when stopped; etc.
Similarly, some techniques harness dark patterns\footnote{The legal definition in California is ``A user interface is a dark pattern if the interface has the effect of substantially subverting or impairing user autonomy, decision-making, or choice. A business’s intent in designing the interface is not determinative in whether the user interface is a dark pattern, but a factor to be considered." CPRA § 7004 (c)}~\cite{gray_mapping_2023} to manipulate users into taking actions or decisions they wouldn't take otherwise.
This is typically the case when one accepts all notifications of an application without really noticing it, while deactivating notifications would require an additional, less intuitive, series of actions.

Secondly, recent advances in machine learning allow the training of \textbf{content recommendation algorithms} on massive online behavioral data, which Zuboff refers to as ``behavioral surplus''~\cite{zuboff2019age}. These algorithms learn to recommend content that not only captures attention but also increases user engagement\footnote{There are multiple definitions of user engagement. In the context of social media, this typically refers to the fact that a user would interact with a content: e.g. like, comment or repost it. Engagement is usually public in that it leaves public traces on the platform, unlike sheer content consumption that remains private~\cite{robertson_negativity_2023}.}.
They discover the content's key features that help predict whether such content will effectively attract users' attention, and typically end up selecting content related to conflictuality, fear or sexuality~\cite{bronner2021}.
They also learn to exploit humans' negativity bias~\cite{soroka_cross-national_2019,siegrist_better_2001} and, as a consequence, content conveying high-arousal negative emotions (such as anger, resentment, indignation and disgust) are more likely to be read and eventually shared online than those conveying other emotions~\cite{robertson_negativity_2023,kohout_may_2023}. 
Concerningly enough, false information (a broad term including misinformation and other forms of disinformation) typically relies on such negative emotions as a trick to foster sharing.
Finally, recommendation systems may do all this without it being explicit, neither in the features they select nor in the succinct feedback that some of them happen to provide users with\footnote{For instance, a recommendation system may tell us ``you liked this movie, you may also like this one''. But we don't know what features were selected to recommend this one: Do they have an actor in common? Did my contacts like both of them? etc.}.

Since the amount of attention available is both limited and precious, it would be unsustainable for a civilization to waste it massively and with impunity for questionable or futile purposes~\cite{bronner2021}.
Today, we might precisely be at that moment: while mental time has become a new oil, we have created an attention economy and subsequent attention markets~\cite{hefti_economics_2015,hendricks_attention_2019} that, although sustainable from an economic point of view, may be unsustainable from a civilization point of view.
From these first references, let us define what the term ``attention market'' refers to in this article.

\begin{definition}[Attention Market]
  Economic environment where businesses compete to capture and retain the resource represented by people's focused mental engagement that we call attention.
\end{definition}

The attention market treats attention as a tradable commodity and involves multiple actors: from producers (the end users whose attention is the resource), to content creators whose work is used to capture the attention, brokers who trade and monetize the attention, and consumers (the large Web platforms) who use it for their own purposes such as exposing users to advertisements.

\sepline

In this article, we propose a discussion aimed at spurring introspection and debate, primarily within the computer science community which designs these technologies, but also more generally within all the communities concerned about societal issues of AI technologies.
Our goal is not to provide an additional piece of evidence, for there are many as we will see, but to insist on the interdisciplinary recognition of the situation, of its urgency and of the possible means of action.
In line with the Web Science Manifesto~\cite{berendt:hal-03189474} calling for interdisciplinary approaches to prepare the future of the Web, we bring together and synthesize the conclusions of more than 70 papers and books from a wide range of disciplines to analyze the practices and drifts of these systems designed to capture attention on a worldwide scale.
We make the point that, with the initial commonplace goal of making targeted advertising more effective, the generalization of attention-capturing techniques and their use of negative emotions tends to foster radicalization and polarization, amplify the dissemination of false information, spur the emergence of populism, and eventually put a threat on democracies and human societies in general.


If we look specifically at the computer science community, it appears that, so far, the public and private research has invested large efforts in dealing with some aspects of the problem like radicalization, violent speech and false information.
Most of these works rely on \textit{post hoc} measures (measures taken after the problems have appeared) such as content detection, deletion or downgrading (decreasing the content ranking so that it is less likely to be recommended).
Nevertheless, we argue that additional measures must be considered to actively prevent the issues that stem from attention capturing rather than only mitigating their impact once they have occurred.
Presumably, such measures would be political and economical as well as technical, meaning that this socio-technical problematic situation calls for socio-technical solutions.
Hence this call for an inter-disciplinary approach meant to regulate the attention market.

In the rest of this paper we will first review the general principles of recommendation systems and the consequences of the recommendation loop that they implement (section~\ref{sec:recommenders}). Then, we will explain how having recommendation systems harness emotions can lead to detrimental situations including what we shall name an algorithmic emotional governance (section~\ref{sec:emotions}). We will touch upon the threat to creative jobs (section~\ref{sec:scientists}) and then review some known post hoc measures (section~\ref{posthoc-measures}), before discussing preventive measures to reclaim our attention (section~\ref{sec:redirect}).


\section{Users in the Loop... of Recommendation Systems}\label{sec:recommenders}

Content recommendation algorithms are a key component of a wide range of applications, including social media, search engines and major Web platforms in general.
Through many applications they have changed our lives, helping us to be more efficient, assisting us in daily tasks, or improving our education and information.
In a number of other applications however, the reality in not so bright.
In the case of social media for instance, they are presented to us as if designed to provide us with content that matches our needs and desires, while what they really seek is to maximize the attention we pay to their hosting platform and advertisements thereof. 

Through the training process, recommendation algorithms automatically learn to extract from massive behavioral traces the content's features that most effectively capture our attention and maximize our continuous engagement with the platform.
Typically, they can learn that some categories of topics, such as conflictuality, fear or sexuality, irresistibly attract our attention~\cite{bronner2021}, and thus lean toward recommending these particular categories.
They can also learn to select content tailored for a certain user, by taking into account both the content's features (topics, source, emotions conveyed...) and its adequacy with the user's profile (interests, inclinations, past behavior...). 
This adequacy likely involves many other features that are not transparent since the platforms rarely inform users about how and for which purpose their personalized feed was composed.
This is underlined in a study by DeVito~\cite{devito2017editors} who analyzed Facebook’s patents, press releases, and Securities and Exchange Commission filings, to identify ``the set of algorithmic values that drive the News Feed''.
Some of the features he identified are objective, i.e. they can be observed or measured: friend relationships, explicitly expressed user interests, prior engagement, post age and page relationships.
By contrast, other features are up to interpretation and thus raise multiple questions: implicitly expressed user preferences (what are the signals of such implicit expression?), platform priorities (what are they and who decides them?), content quality (what are the quality criterion?).

Finally, it may seem that recommendation algorithms learn to leverage psychological traits and cognitive biases. Yet, it is important to stress that the algorithm does not discover such things as a psychological trait or a cognitive bias itself.
Rather, it discovers the features that enable it to exploit what psychologists would describe as a trait or bias.
Such criteria are not explicitly formulated, they may not even be explicable nor verifiable. They remain implicit in the models unless a study be carried out a posteriori, that would surface the biases that emerge from the recommendations.
This is yet another example where AI techniques, without explanations nor feedback, raise ethical concerns.

Another specificity of recommendation algorithms is that they tend to implement a self-reinforcing loop that we define as follows:

\begin{definition}[Self-Reinforcing Recommendation Loop]\label{def:rec_loop}
  The continuous cycle of recommendation systems providing personalized suggestions to a user based on data collected from their preferences and behaviors and integrating these to further recommendations.
\end{definition}

A classical self-reinforcing recommendation loop is illustrated in figure~\ref{fig:loop}:
\circled{1}~The algorithms recommend content to the user.
\circled{2}~The behavior of the user is captured, possibly partially due to the focus of the platform and the limited choices that the interface offers, and possibly biased due to the fact that these choices may be oriented, again by the interest of the platform and the chosen interface. 
\circled{3}~The algorithms integrate these reactions in future recommendations. As a result, the reactions of the user will reinforce the recommendation and propagation of the content.

\begin{figure}
    \centering
    \begin{tikzpicture}[
    node distance = 5mm and 5mm,
      start chain = going right,
       arr/.style = {-{Straight Barb[angle=60:2pt 3]}},
       box/.style = {draw, font=\small, align=center, minimum height=6mm, on chain, join},
      bbox/.style = {draw, rounded corners=3pt, font=\small, align=center, minimum height=6mm, fill=black,text=white},
      boxr/.style = {box, rounded corners=3pt, fill=black!15},
     ellip/.style = {ellipse,font=\small, draw, minimum width=3em, minimum height=6mm, inner xsep=-2pt, inner ysep=1pt, align=center, on chain, join},
     every join/.style = {arr}   ]                            
    \node (n1) [ellip]  {\Gentsroom~Connect};
    \node (n2) [boxr]   {\Industry~Select (1)};
    \node (n3) [ellip]  {\Gentsroom~Perceive \& React};
    \node (n5) [boxr] [below = of n3]  {\Industry~Trace (2)};
    \node (n6) [bbox] [below = of n2]  {\Industry~A.I. processing (3)};
    
    \draw [->] (n5) to  (n6);
    \draw [->] (n6) to  (n2);
    
    \end{tikzpicture}
    \caption{The self-reinforcing recommendation loop of platforms: the ellipses are activities on the user side, the boxes are activities on the platform side. \textit{Select} and \textit{Trace} are grey boxes because only partially observable. The A.I. processing is, more than often, a black box for the end-user.}
    \label{fig:loop}
\end{figure}


Of course there are externalities to that loop, that can increase its impact.
Smartphones, for example, provide additional means to profile users by tracking their every moves, making recommendation even more efficient and targeted to the point that it competes and sometimes takes over more traditional ways of advertising~\cite{the_economic_times_no_2017}.
Another (detrimental) externality of this loop is that it opens the door to spoofing techniques and other malevolent actors intentionally biasing usage traces to ``hijack'' recommendation systems. Indeed, as soon as a process is known and documented, it runs the risk of being diverted from its original purpose and manipulated beyond its original objectives. 
For instance, fake reviews and reactions alter recommendations; black hat techniques of SEO (Search Engine Optimization) such as hidden texts, link farms, cloaking\footnote{Cloaking denotes a technique in which the content presented to a search engine crawler is different from that presented to an actual user. It aims at deceiving search engines so they display the page that they would otherwise downgrade or dismiss. Adapted from \url{https://en.wikipedia.org/wiki/Cloaking}.} or text spinning are disapproved by search engines as they impact the recommendations they make by unduly increasing the ranking of targeted pages or avoiding their downgrading.

\sepline

As a result, the fact that a few recommendation systems influence a significant fraction of the human population may have a number of detrimental side effects on their users and our societies at large.
A first side effect is that recommendation algorithms tend to lock users in an informational space in accordance with their tastes and beliefs, a ``filter bubble''~\cite{pariser2011filter} that confines them to a ``cognitive comfort zone'' and activates their confirmation bias as they are faced with information which seems to go towards the same directions or conclusions~\cite{sasahara_social_2021,kitchens_understanding_2020}.
Eventually, users are no longer confronted with contradiction, debate nor disturbing facts or ideas, and
this algorithmic amplification tends to be a powerful driver of the radicalization and polarization of opinions, leading to extremist ideas in some cases~\cite{whittaker_recommender_2021}.

Furthermore, at a time where we need to change our behaviors (e.g. over-consumption of goods and energy) and redirect our attention to important matters (e.g. climate change), we should question whether recommendation algorithms make the right recommendations, and for whom.
Considering the billions of users caught in recommendation loops everyday\footnote{In 2018, Google revealed that 70\% of the time spent watching videos on Youtube is about videos recommended by Youtube's algorithms. \url{https://qz.com/1178125/youtubes-recommendations-drive-70-of-what-we-watch}}, it is important to continuously monitor how and for what purpose these systems capture our attention. Because when our attention is spent on a content chosen by these platforms, it is lost for anything else.


\section{Algorithmic Emotional Governance}
\label{sec:emotions}

Considering the platform's recommendation loop introduced in section \ref{sec:recommenders}, we now want to stress that, directly or indirectly, emotions are a key feature of the selected recommendations. In fact, the whole attention market could be seen as driven by a complex equation involving, at least, emotions, cognitive biases and content recommendation algorithms. 
This could lead to what Patino calls an ``emocraty, a regime that makes our emotions become performative and invade public space''~\cite{patino_lemergence_2024}.
Here we will rather speak of an \textit{algorithmic emotional governance} merging two concepts: emotional governance~\cite{richards2007emotional} which is the informed management of the emotional dynamics of the governed population,
and algorithmic governance~\cite{rouvroy2013gouvernementalite} which is a governance of our societies based on the algorithmic processing of massive data. 

\begin{definition}[Algorithmic Emotional Governance]
The governance of societies based on algorithms processing massive data to harness the emotional dynamics of the governed population.
\end{definition}

Emotions are a powerful attractor of our attention, 
especially emotions with a high negative valence~\cite{soroka_cross-national_2019}. As a result, information that arouses anger, fear, indignation, resentment, frustration or disgust is among those that most effectively catch our attention~\cite{robertson_negativity_2023,kohout_may_2023}.
An explanation is that witnessing others' negative emotions activates our comparison bias and subjects us to some sort of injunction to take sides, to show our emotional response, and hence publicly demonstrate our ``irreproachable morality''~\cite{crockett_moral_2017,bronner2021}.
Note that catching attention and increasing user engagement are different things, and although high-arousal negative emotions catch attention more efficiently than other emotions, it remains unclear whether they induce a higher user engagement on social media. In some cases a higher sharing rate of information conveying positive emotions was observed~\cite{keib_picture_2018,kramer_spread_2012}. 
Nevertheless, in several contributions,
researchers showed the overwhelming impact of emotions in argumentation and debates and the means to detect them ~\cite{BenlamineCVCFG15,VillataBCFG18,BasileCVFG16}, and it has also been shown that anger spreads faster on social media than any other emotion~\cite{fan_anger_2014}.
Note that this attraction for negative content can be observed in completely different domains, e.g. in literature where the anti-utopian and dystopian fiction genres became more prominent within the utopian genre~\cite{mascarell2020bibliometric}.

Combined together, the construction of filter bubbles by recommendation systems and the ability of these systems to learn the content's features that trigger a particular emotional response in a particular individual, can lead to some form of polarisation and end up trapping users in radicalization pathways.
Consider the supporter of a sports club: it is because the system chooses the right topic (e.g. the right sport), the right content (e.g. an article about an opponent club) and the right tone and emotion (e.g. mocking criticism) that an emotion is provoked, followed by a registered reaction (like, comment, repost) and, over time, a potential polarisation is developed such as hatred for the opponent's supporters.

\begin{figure}[h]
\centering
\fboxsep=1pt
\fboxrule=0.5pt
\fcolorbox{black}{white}{\includegraphics[width=0.465\textwidth]{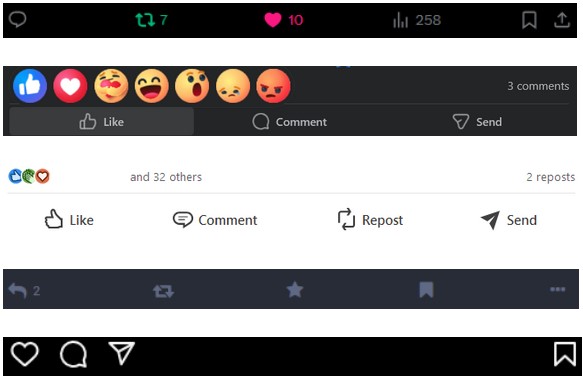}}
\caption{Five interfaces of well-known social media to react to a post.}
\label{fig:five-interfaces}
\end{figure}

Recommendation after recommendation, the filter bubble becomes an opinion bubble where users are isolated from discrepant opinions, and eventually an emotion bubble where they are maintained in certain emotional states. 
In the end, the complex interaction of negative emotions, cognitive biases (e.g. negativity bias and impulsive tendency to show indignation) and recommendation algorithms leads to an emotional escalation.
Often, this escalation is further worsened by the affordances offered by the platforms, that tend to make exchanges ever briefer and more simplistic: 
How to express a nuanced reflection in a 280-character tweet? How to underline a doubt when the only available choices are essentially limited to~\faHeartO/\faThumbsOUp~(and sometimes \faThumbsODown)? 
How to agree with one part of a post and disagree with another one when this post is treated as a monolithic block by the interface that only offers the options \faRetweet/\faShare/\faShareAlt~?
Some examples of widgets from well-known platforms are illustrated in figure~\ref{fig:five-interfaces}.
This extreme discretization of choices adds to the mechanisms at work and reinforces the polarization of opinions and communities.
Some dark patterns are even intentionally employed to make some actions easy and some more difficult: for instance, in Facebook the button to like a post is always visible whereas the option to report a post is at the bottom of a sub-menu, a pattern falling in the category known as ``longer than necessary''~\cite{edpb_guidelines_2022}.

Eventually, nuance, doubt or agnosticism are mechanically made invisible because the low emotional response that they induce simply downgrades their ranking. It is imperative to have an opinion, preferably definite and cleaving.
Amplified by digital disinhibition\footnote{the feeling of impunity induced by the feeling of anonymity}~\cite{suler_online_2004}, this emotional escalation can lead to outpourings of violence and hatred whose outcome is sometimes tragic as attested by the suicide attempts of teenagers being cyberbullied~\cite{schonfeld_cyberbullying_2023}.
Moreover, the full consequences of triggering or regulating emotions on our cognitive functions in general and on memory in particular remain to be studied extensively~\cite{richards2000emotion}.

\sepline

We just described the combined effect of emotions, cognitive biases and recommendation algorithms, which is at work whatever the type of content a platform serves.
But things get even worse when it comes specifically to false information.
False information are frequently meant to arouse strong negative emotions~\cite{zaeem_sentiment_2020}, and the combination with cognitive biases and recommendation algorithms provides them with a particularly fertile ground and a formidable cognitive efficiency~\cite{acerbi_cognitive_2019,martel_reliance_2020}.
Some studies reported that negatively biased fake news enhance users' willingness to share them~\cite{corbu_fake_2021}, and reveal a positive correlation between the virality of fake news and the anger they carry~\cite{chuai_anger_2022}.
Another study contended that falsehood spreads ``significantly farther, faster, deeper, and more broadly than the truth'' on social media~\cite{vosoughi_spread_2018}, which underlines that recommendation of content arousing negative emotions does not only induce local individual reaction: it creates a chain reaction leveraging the network effect of social media to spread that ``content--emotion'' couple through the acquaintance links.
Other studies reported that recommendation algorithms mechanically tend to favor false information conveying divisive ideas, shocking events and negative emotions~\cite{fernandez_analysing_2021,fortunato_social_2022}.
This type of content entails a felt injunction to take sides and compulsively spread shocking information rather than appeal to critical thinking, question its veracity and verify its source.
And since this information is often relayed by acquaintances, the social proof bias~\cite{Cialdini_2008} entices users to deem it credible and trustworthy.

Concerningly, the contents we are exposed to leave a trace in our implicit memory: although we cannot recall seeing it, it may impact our choices for several months~\cite{jcc4.12035}.
Even more concerning is the fact that, due to the negativity bias, negative information leaves a longer memory trace than positive information. Therefore, even when a false information is denied or rectified, there remains a negative feeling that stems from the strong emotional response it triggered in the first place.
Repeated again and again, associated with representations of the world that summon conspiracy theories, reinforced under the pressure of filter and emotion bubbles, and propelled by the network effect, such information tends to gradually and insidiously undermine our trust in experts (scholars, journalists, etc.), entails risks for public health~\cite{waszak_spread_2018,rocha_impact_2023}, and spurs the emergence of extreme ideas and populism that eventually undermine democracies~\cite{whittaker_recommender_2021,abreu_algorithm-driven_2021,intelligence_social_2022,fortunato_social_2022}, among other pitfalls.

Finally, let us stress that if ``previous studies have shown how personality, values, emotions and vulnerability of users affect their likelihood to propagate misinformation''~\cite{fernandez_analysing_2021}, in this section we only considered an average user without any particular health condition. 
But we should envisage more complex situations when it comes to users with disabilities or mental disorders e.g. depression, anxiety, compulsive shopping disorder, paranoia, FOMO, FOBO\footnote{Fear of Better Options: the inability to choose when faced with a multitude of options. 
}...
Let us just mention one specific condition: the attention deficit (AD) disorder. There is evidence that AD symptoms could be worsened by the use of digital media and their attention-grabbing applications, and more importantly that these applications could provoke AD among people without previous record of such a disorder~\cite{ra_association_2018}. To the very least, more research is needed in this respect.

\section{Attention, Attention, all Thinkers}\label{sec:scientists}
We firstly intended this section for all the scientists reading this paper, concurring with the article of David R. Smith: ``Attention, attention: your most valuable scientific assets are under attack''~\cite{smith_attention_2018}. In this article, Smith calls for attention to what media platforms are doing to research and the academic domain.
Indeed, even the most informed scientists and engineers are not immune to these problems~\cite{lewis_our_2017} such that digital contraptions (as Smith calls them) are contributing to \textit{academic attention deficit disorder}~\cite{smith_attention_2018}.
In fact, concentration but also boredom, mind-wandering and daydreaming times are vital to creative thinking. Many of us experienced the sudden burst of an idea in the middle of a relaxing moment. Attention-capturing systems 
steal these moments from all of us 
and hamper the creativity process of wondering minds~\cite{zomorodi2017bored}.

Of course these remarks can be generalized to many other activities requiring concentration, creativity and imagination, and one could wonder what digital contraptions are doing to politics, healthcare or education, for instance. 
To mention just one example, countless information media now report the cases of Youtubers experiencing a burnout~\cite{parkin_youtube_2018}, or musicians complaining that they spend more time making Tiktok videos to promote their music than actually creating music~\cite{shah_making_2022,whateley_tiktoks_2023}.
This reveals that, to hook and keep the attention of content consumers, platforms also exercise some sort of visibility tyranny over content creators.

In other words: attention, attention, thinkers, we need to redesign the systems for our own needs, rather than the other way around, 
especially in creative jobs since the true currency of these jobs are ideas~\cite{smith_attention_2018}.


\section{Known Post Hoc Counter-Measures}\label{posthoc-measures}

Among the various issues raised in the previous sections, the questions of false information, radicalization, hateful speech and bullying are among the most concerning, and therefore have been extensively addressed by the research community~\cite{sharma_combating_2019}.
In~\cite{fernandez_analysing_2021} authors identified three different points where recommendation systems can be adapted to tackle these issues: \circled{1}~pre recommendation, \circled{2}~within the recommendation model, and \circled{3}~post recommendation. 
Most of the current counter-measures to deal with false information lie in this third category. Below we touch upon some of them.

Firstly, to dyke the spread of false information as well as inappropriate content such as bullying, hateful or violent speech, social media and content hosting platforms have obligations that vary depending on the legislation and its jurisdiction~\cite{funke_guide_nodate}.
Measures range from content deletion and suspension of users spreading inappropriate content, to re-ranking of recommended items before presenting them to the user~\cite{fernandez_analysing_2021}, flagging to indicate potentially deceptive content, etc.
Yet, despite these various approaches,
progress is still necessary.
For instance, subtle violent content may be hard to detect as soon as it does not contain explicit hateful or violent terms, or when it uses sarcasm~\cite{ocampo-etal-2023-depth}.
Conversely, content may be erroneously assessed as abusive or illicit although it is in fact using irony to convey perfectly acceptable ideas. An in-depth analysis of implicit and subtlety in linguistic content remains an open question~\cite{ocampo-etal-2023-depth}.

In addition, any action must carefully consider the dangers of transferring regulation and enforcement to private companies.
~\cite{stark2020algorithms} argue that over-filtering content is just as dangerous as letting bad content spread.
Indeed, deletion and filtering may deviate from initial purpose to over-censorship of content if it becomes safer for the platforms to do so than take a risk of being sued.
Furthermore, assessing the trustworthiness of information raises multiple ethical and political concerns: Who decides what is true or false? According to which criteria? Under whose control? 

Secondly, to mitigate the effect of false information, multiple post hoc measures rely on the impact of additional corrective content. 
For~\cite{vraga2020correction}, pointing to a coherent alternative explanation, with references to expert and highly credible factual sources, remains a solid starting point.
The authors describe the strategy of ``observational correction'' leveraging the fact that users who witness the correction of a misinformation item, but have not directly engaged with that item, are less affected by cognitive dissonance and are therefore more amenable to correction.
This is consistent with the findings of~\cite{bode2015erratum} who suggest that exposing users to related stories that correct a post that contained misinformation will significantly reduce misperceptions. 
The impact of the correction can be further reinforced by explicitly pointing to the demographic similarity between the user and the authors of opposing content~\cite{garrett2013undermining}, which taps into the homophily effect\footnote{Homophily: the tendency to associate with similar others.}.
In other words, we are more likely to accept the correction when it comes from someone who is socially close to us,  e.g. having the same professional activity or background.
\cite{vraga2020correction} also suggest to multiply correction actions for each targeted content to reinforce the effect.


\section{Reclaiming Our Attention}
\label{sec:redirect}

The methods presented in the previous section all have one thing in common: they deal with the problems in a post hoc manner, that is, after these problems have occurred, with all the limitations that come with this ``coming after''.
To go further however, we need to figure out measures, may they be legal, economical or technical, capable of preventing the attention from being looted in the first place.
More importantly, we need to consider this reflection not only from the perspective of regulating the attention consumers (the platforms and multiple intermediaries), but also from the perspective of the producers (the end-users) who want to reclaim their attention, especially in times when our attention is needed on a number of urgent matters.
This involves actively preventing recommendation systems from finding ways to exploit our inner limitations and manipulate us through sometime ancient and deeply embedded structures of our brain (e.g. our striatum)~\cite{bohler2019bug}.

Below we formulate a set of propositions stemming from the observations and findings reported in the previous sections. We organize them  around the challenges that they address, together with suggestions made by other authors from multiple disciplines. Finally we extract from them a set of empirical principles that could be used do drive further works on good practices.


\subsection{The Carrot and the Stick}

Taking the example of false information, \cite{vraga2020correction} insist on the fact that a posteriori corrections are not sufficient and must happen as early as possible, that is, before misperceptions are entrenched.
Besides, avoiding the algorithmic amplification effect of such information by recommendation systems requires to mitigate the popularity effect before its happens~\cite{fernandez_analysing_2021}.
But if online social media are required by law to combat false information, they have conflicting incentive to do it, not to say no incentive at all. 
Indeed, as we described in section~\ref{sec:emotions}, false information largely relies on negative emotions to capture users' attention. As such, they are very effective in fostering user engagement which is what online social media strive to obtain.
Consequently, from an economic point of view, it is counter-productive for them to prevent the spread of false information.
More generally, it is counter-productive for platforms to mitigate the popularity effect, mitigate the impact of negative emotions, or reduce filter bubbles and the subsequent polarization of opinions.

The authors of \cite{newman_regulating_2019} suggest to rethink existing trade regulation laws such as antitrust and fair competition laws under the new realm of attention markets. 
They propose to enforce taxes on attention consumption to ``disincentivize attention intermediaries from vacuuming up as much attention as possible'', for instance by restraining the amount of advertisements that can be shown to a user, or reducing the deductibility of advertising expenditures from the companies' revenues to alleviate their taxes.
They also propose to regulate the attention costs that can be charged, with the idea that if attention becomes less lucrative then financial resources will be redirected towards more lucrative markets, thus reducing the amount of attention being captured and traded.
Such regulation could take the form of a tax on the daily attention time being caught beyond a certain threshold. Another tax could be enforced on the individual and collective harms caused by attention capturing techniques, which first requires to explicit those harms like it has been done for dark patterns~\cite{santos_no_2024} and more precisely to qualify~\cite{beckers2022causal} and quantity~\cite{beckers2023quantifying} the harms done.

In other words, things would not change without strong incentives on one side, and disincentives on the other. We can summarize this in the following general principle:

\begin{principle}{\faThumbTack ~ principle of the right incentive}
Governance bodies should leverage legal and economic means to drive platforms' practices towards desirable behaviours, while penalizing undesired behaviours.
\end{principle}

\subsection{Usage Regulation}
Some of the measures meant to regulate the attention market lie in the way the services are consumed.
As has already been done in some countries, laws could be voted to limit the daily time spent by users on certain services, especially among the youngest~\cite{Kantrowitz_5_2021,carroll_much_2023}.
Another simple measure applies to video streaming platforms, that consists in imposing few-second pauses between videos. This apparently naive technique may actually shatter the infinite feed trap by giving users the short amount of time they need to realize that they have been in an attention tunnel for a while, and that they want to ``reclaim'' their attention.
This can be generalized by formulating the following principle:

\begin{principle}{\faThumbTack ~ principle of supported due diligence}
All means should be provided to foster and update the due diligence of users. In particular they should always be given and made aware of the options to escape the systems' loops, processes and goals.
\end{principle} 

Policy makers could also tackle the problem of attention fragmentation entailed by the multiple, often invasive, notifications that smartphone applications raise. Whenever a notification occurs, users are tempted to interrupt their current activity, check the reason of the notification, possibly react to it, before eventually returning to their activity.
It has been shown that switching our attention between tasks or contexts
has a cost: it is time-consuming and creates a more error prone context~\cite{jersild1927mental,monsell2003task,rogers1995costs}.
Furthermore it has even been shown that the mere presence of such devices, although turned off, impairs our cognitive capacity~\cite{ward2017brain}. 
In a mindset similar to the European General Data Protection Regulation (GDPR)\footnote{General Data Protection Regulation (GDPR) \\  \url{https://gdpr-info.eu/}}, which imposes the consent of users for the use of cookies,
the EU Parliament has recently adopted a resolution on addictive design to put forward ``a digital ‘right not to be disturbed’ to empower consumers by turning all attention-seeking features off by design''~\cite{european_parliament_addictive}.
This could lead to laws imposing that smartphone applications obtain users' explicit and informed consent for the notifications that they raise, and deactivate them by default (``opt-in only'').
Hence the following principle:

\begin{principle}{\faThumbTack ~ principle of opt-in by default}
Recommendation and notification services should be turned off by default and only turned on on demand and after informed consent and preference setting
\end{principle} 

This could be complemented by more punitive measures, as proposed in \cite{mudde2017populism}, for instance by demonetizing and forbidding collaboration with platforms that do not follow the rules.


\subsection{Content Recommendation Monitoring}
The echo-chamber effect of recommendation algorithms is at the root of multiple examples of polarization and radicalization. It could be mitigated by imposing a certain share of non-recommended content, content that is outside of the user's interests, or content that originates from users they are not acquainted with.
In this respect, some approaches lie in the second category proposed by ~\cite{fernandez_analysing_2021}, i.e. modifications ``within the recommendation'' system. 
For instance, the same authors suggest using clustering approaches to assemble the contacts of the user according to different levels of similarity with the user, and leverage these groups to increase the diversity of recommendations while maintaining a certain coherence and similarity.
\cite{fabbri_rewiring_2022} propose a method to come up with relevant recommendations while reducing the likelihood of enticing the user towards radicalization pathways.
Also, to counter the misuse and abuse of anger, indignation or fear, which are often associated with false information, platforms could be required to carry out sentiment analysis on every content in order to keep the amount of recommendations associated to negative emotions below a given threshold.

\begin{principle}{\faThumbTack ~ principle of balanced recommendations}
Recommendation-based platforms should prevent the over specialization of recommendations w.r.t. all features and should support monitoring and prevention of the formation of bubbles of any type (opinion, source, emotion, etc.).
\end{principle} 

Moreover, there exists an asymmetry of visibility between the viral spreading of an information that was proven to be false or misleading, and the denial or rectification of that information.
The denial of a false information usually puts forward a pondered, nuanced position that appeals to reasoning and facts (\textit{logos}) over emotion (\textit{pathos}). Hence, it does not trigger an emotional response compared to the one generated by the false information in the first place, and it is therefore silently downgraded by recommendation algorithms. This is commonly summarized by the so-called Bradolini's law which states  that ``the amount of energy needed to refute false information is an order of magnitude bigger than that needed to produce it.''
As a result, users who propagate false information often never get to know about their mistake.
~\cite{venturini2019fake} insist on the fact that it is critical to jointly address content-checking and digital virality. Thus, to counterbalance this visibility asymmetry, social media could be required to impose on the denial/rectification of a false information a visibility equivalent to that of the initial information, for instance, by ensuring that the population who was exposed to the false information be exposed to the denial too.
A warning could also be presented to users who propagated this false information in order to increase their awareness.
Of course, this type of measure could be coupled with other post recommendation measures such as strategies involving the observational correction or demographic similarity presented in section~\ref{posthoc-measures}.
More generally, one has to figure out how we can use recommendation systems to recommend counter measures, \textit{i.e.} we could train a recommendation system to learn the most relevant content and the most impactful entries in the acquaintance network to inject a counter measure.

\begin{principle}{\faThumbTack ~ principle of balanced visibility}
Recommendation-based platforms should ensure that preventive and corrective measures have a visibility at least equal to the visibility of the problems being prevented or corrected.
\end{principle}


\subsection{Affordances and Interaction Design}
As discussed in section~\ref{sec:emotions}, the frictionelss affordances of platforms are optimized towards extremely brief and basic exchanges, to favor their quantity over their quality, leaving no room for nuance, pondering, doubt nor substantiated reasoning.
Interfaces could be redesigned to facilitate non-binary reactions, starting with a range of nuanced emotions. 
Rather than implementing deceptive dark patterns, they could restore ``desirable frictions''~\cite{broadbent:hal-04542396} to prevent compulsive, consciousless behaviors, and rely on nudges to gently drive users towards critical thinking, and by valuing/rewarding this kind of behavior.
~\cite{abreu_algorithm-driven_2021} recommend engaging users in the validation of content before sharing it, both manually and with automated analysis methods on content and context.
For instance, X (formerly Twitter) asks confirmation before retweeting the link to an article that the user did not click. 
Similarly, interfaces could encourage users to comment on content instead of merely clicking \faHeartO,~\faThumbsOUp~or~\faThumbsODown, and they could question a user about whether they really want to share or support a content associated to strong negative emotions or for which a counter-measure was triggered.

\begin{principle}{\faThumbTack ~ principle of benevolent interaction design}
Affordances and interactions should be designed and evaluated with the well-being of end-users in mind first.
\end{principle}


\subsection{Societal Impact and Educational Mission}

We, as a society, could decide that large online social media, because of their influence on the society, public opinion, public health and economy, can no longer be considered as sheer private companies regulated by markets law only. Instead, they could be seen as digital commons and be assigned a specific status that would endow them with a societal mission including, for instance, an educational purpose. 
As an example, they could instruct users in detecting false or misleading information, 
they could promote content meant to increase awareness w.r.t. attention mechanisms and cognitive biases, foster assertiveness, critical thinking and ``distill'' concepts of the scientific method, etc.
On the same page, authors of~\cite{abreu_algorithm-driven_2021} insist on the need for civic education, and~\cite{oneil2017weapons} recommend integrating democratic values into the algorithms that impact our lives, especially the ones participating in an algorithmic governance (e.g. platform used for debates, for information, for legal actions, education).
States and civil societies could favor the entrance of new platforms that, instead of optimizing user engagement and attention catching, would favor democratic debate in the digital space by supporting free constructive debate and the collective development of solutions to major societal issues.
An illustration of this approach is that of Polis, an open-source social network meant to support large-scale public debate~\cite{small_polis_2021}. Its recommendation algorithm is trained to promote consensus.
During experimentations, particularly in Taiwan, Polis has ``shown itself capable of 
building shared understanding, disincentivizing counterproductive behavior (trolling), and 
cultivating points of consensus''.

\begin{principle}{\faThumbTack ~ principle of digital commons preservation}
When a digital service, platform or resource reaches the potential of having a world-wide impact on human societies, it must be assigned the status of digital common and must be subjected to preservation rules and policies.
\end{principle}


\subsection{Feedback and Transparency Enforcement}
One of the pitfalls we identified is the fact that users are not aware of being caught by recommendation loops. Furthermore, it is established that online sharing of fake news increases with social media fatigue~\cite{talwar_why_2019}.
Therefore, approaches such as \textit{quantified self} and \textit{lifelog} could be specialized to the case of recommendation-based platforms, in order to foster awareness and introspection.
Some online services already provide their users with usage metrics with respect to the total time spent on the platform, or reminders to make a break at regular intervals.
This can be seen as a shy improvement towards compliance with the European Digital Services Act~\cite{european_union_regulation_nodate}, that took effect in August 2023, however such options remain deactivated by default, and usually accessible from impractical sub-menus. 
Self-tracking tools could provide users with a much larger set of metrics, indicators and feedback such as the total exposure to negative news, the diversity of the recommendations they are shown, and make them aware of low-diversity risks. For instance the fact that ``90\% of the content one sees come from 10\% of one's contacts or are on the same topic'' may indicate that one is experiencing a filter bubble.
In other words, platforms must set up the conditions for individual and collective reflexivity, that is, the ability of users and communities to be self-aware of their usage and engagement with the system. 
Hence the following principle:

\begin{principle}{\faThumbTack ~ principle of continuous reflexivity}
Users must be provided a continuously updated feedback on their usage of the system and on themselves to support their reflexivity and maintain an up-to-date informed consent.
\end{principle}

This reflexivity requires not only usage metrics but also transparency about the platform's purposes.
Among other measures, the European Digital Services Act requires that platforms set up mechanisms to explain the reasons that led to recommending a certain content, and to offer users an alternative recommendation not based on profiling.
Such measures are especially crucial when coupling AI and the Web since we need to set transparency and explanation as a prerequisite to any approach, to ensure the awareness and informed consent of, potentially, billions of users \cite{berendt:hal-03189474}.

\begin{principle}{\faThumbTack ~ principle of full user awareness}
Users must be made aware of all the features and purposes leading to a recommendation, before and when it is provided.
\end{principle}

Let us add that, for the regulations proposed above to be enforced effectively, one cannot rely solely on the goodwill of online platforms.
But given the scale of these platforms, public actors have neither the information nor the technical possibilities to evaluate the application and effectiveness of the regulations~\cite{zolynski_leconomie_2019}. Hence the following principle:

\begin{principle}{\faThumbTack ~ principle of observability}
Online platforms must provide institutions, civil society and researchers with the legal and technical instruments allowing them to carry out active control and verification of the application and effectiveness of regulations.
\end{principle}

To the very least, this requires open and free access to platforms data by researchers, as opposed to the recent decision of X to deny them free API access.


\subsection{Build on Existing Practices}
Finally, and although it may seem obvious, one rule is worth remembering: to review and take inspiration from existing best practices in other domains.
In most jurisdictions there exist advertising laws to protect consumers, ensure they remain able to make informed decisions, and more generally to maintain a level playing field\footnote{Metaphor denoting the fact that, in business, all players compete fairly, i.e. they all play by the same set of rules. \url{https://en.wikipedia.org/wiki/Level_playing_field}} between all players. 
Most countries regulate advertising through legislations that target different forms of false, misleading or deceptive advertising contents and claims, and forbid a whole range of practices (unsubstantiated comparison, forged testimonial, puffery, misleading packaging/label, unsolicited commercial messages, alleged contests and sweepstakes, etc.). The work and literature on regulating advertising should be reviewed and built-upon in regulating the attention market at large.
This topic is also close to that of clickbaits that are recommended links designed to attract attention and to entice users to follow them while being typically deceptive, sensationalized, or otherwise misleading. Clickbaits are not just teasers but headlines with an element of dishonesty, ``using enticements that do not accurately reflect the content being delivered''\footnote{Definition adapted from \url{https://en.wikipedia.org/wiki/Clickbait}}.
As far as we know, there is no regulation of clickbait practices on the Web, although some of these techniques bear similarities with the misleading or deceptive advertisement practices that we just mentioned and that, on the contrary, are regulated.

\begin{principle}{\faThumbTack ~ principle of best practice transfer}
Methods and tools used to regulate similar situations in relevant domains should be surveyed, benchmarked and systematically considered as input to a Web and AI governance.
\end{principle}

To give another example coming from a completely different angle, we know that parenting practices in terms of TV viewing have an impact on the behaviour of young watchers~\cite{BARRADAS2007369}. Again, approaches and good practices in this domain, and more generally in educating and parenting in the digital media age~\cite{coyne2017}, must be considered in the case of ``Web viewing'' in general and when addressing the problem of attention capturing in particular.

More generally speaking, we need to put in place a governance bodies, starting with the Web and AI, that are prepared to tackle new problematic practices and regulate them, as is done in other areas of activity. 
And we also need to keep a constant watch on these other areas, if only to draw inspiration from the initiatives and feedback they have on similar issues. 
Taking the example of the video game industry, there is evidence of a relationship between ``loot box''\footnote{``loot boxes'' are video game items with randomized contents that can be paid for with real-world money.} spending and gambling addiction~\cite{zendle2018video}, and that a loot box is psychologically akin to gambling~\cite{drummond2018video} and can result in addictive behaviors and endangered players.
The way to study and address that unwanted exploitation of our behaviors is inspirational for other problematic practices on the Web such as those we surveyed.


\section{Thank you for... your Attention} 
\label{sec:thankyou}

AI is domain-independent. It is being applied in all of our daily life areas of interest: information, business, money, politics, employment, sports, games, sex etc. 
And the worldwide deployment of these techniques, partly due to its coupling with the Web, could have detrimental consequences in all these areas alike, unless properly regulated.
This is a commonplace observation but it is the reason why, to prevent such detrimental effects, an ethical AI approach to AI governance must be multidisciplinary and interdisciplinary.

With this mindset, this paper brought together conclusions from more than 70 articles and books  from different disciplines (psychology, sociology, neuroscience, politics, legal domain, computer science, education, etc.) to analyze and call for actions against the current practices competing to capture our attention in several ``Web Wild West corners''.
The problem is both critical and complex, and authors of~\cite{kitchens_understanding_2020} defend the need for a ``nuanced multidimensional view of how social media use may shape information consumption'' and they urge us to consider ``the complex variety in social media platforms [and the] considerable variation in observed impacts among them''.
In ~\cite{fernandez_analysing_2021} authors add that 
``This research requires to navigate the careful tension between privacy, security, economic interests, censorship and cultural differences, and requires to be addressed from multiple disciplines that can assess not only the technological aspect, but also the individual and the social one (...) There is ample room for investigation (...), opening a novel, exciting and interdisciplinary line of research.''

At the same time, the problem is getting worse with every technological innovation.
The pervasiveness of smartphones in our lives has further reinforced the effectiveness of these techniques that can now grab our attention at every moment of the day, and in particular these moments that were previously those of boredom, waiting, daydreaming or intellectual strolling. 
As we pointed out in section~\ref{sec:scientists}, these moments are known to be necessary to spur imagination and creativity. In the continuation of smartphones, smart objects and the resulting internet of things and Web of things will only make things worse.

Recommendation systems that learn to predict us effectively learn to manipulate us, and to be predictable is to lose freedom. Everyday we fuel the predictors in exchange for immediate satisfaction and instant pleasure, 
this amounts to continually mortgaging our freedom.
Besides, these systems that compete for our attention end up pressuring us to consume and to react more and more quickly to their recommendations. And, as we know, acceleration is a form of alienation~\cite{rosa2017alienation}.

In another context and to address our own human limitations, \cite{bronner2021} recommended to find ways to increase our overall level of consciousness and reclaim the power of long-term reflection. 
Our leaders and role models\footnote{A role model is a person whom others look at as an example to be imitated.} struggle to embody the values of patience, conscience and moderation~\cite{bronner2021}, but our computer systems rarely drive us in that direction.
On the contrary, current AI applications are pushing us not to use our conscience, but to play their automation game.
Yet there is no reason for these systems to live in our mind rent free and it is urgent to redesign them so they regularly push us to take a step back, to be more conscious of what we are doing, viewing, saying, spreading, etc. 
The challenge is to (re)take and (re)give time for awareness, attention and reflection: we need to  (re)claim that source of freedom. And for this, we proposed a non-exhaustive first set of principles to (re)design applications and inscribe in them a set of agreed-upon values.


\bibliography{aaai24}

\begin{thebibliography}{86}
\providecommand{\natexlab}[1]{#1}

\bibitem[{Abreu and França(2021)}]{abreu_algorithm-driven_2021}
Abreu, C.; and França, L.~A. 2021.
\newblock Algorithm-driven populism: {An} introduction [{Populizm} oparty na algorytmach. {Wprowadzenie}.].
\newblock \emph{Archives of Criminology [Archiwum Kryminologii]}.

\bibitem[{Acerbi(2019)}]{acerbi_cognitive_2019}
Acerbi, A. 2019.
\newblock Cognitive attraction and online misinformation.
\newblock \emph{Palgrave Communications}, 5(1): 1--7.

\bibitem[{Barradas et~al.(2007)Barradas, Fulton, Blanck, and Huhman}]{BARRADAS2007369}
Barradas, D.~T.; Fulton, J.~E.; Blanck, H.~M.; and Huhman, M. 2007.
\newblock Parental Influences on Youth Television Viewing.
\newblock \emph{The Journal of Pediatrics}, 151(4): 369--373.e4.

\bibitem[{Basile et~al.(2016)Basile, Cabrio, Villata, Frasson, and Gandon}]{BasileCVFG16}
Basile, V.; Cabrio, E.; Villata, S.; Frasson, C.; and Gandon, F. 2016.
\newblock A Pragma-Semantic Analysis of the Emotion/Sentiment Relation in Debates.
\newblock In Lieto, A.; Bhatt, M.; Oltramari, A.; and Vernon, D., eds., \emph{Proceedings of the 4th International Workshop on Artificial Intelligence and Cognition co-located with the Joint Multi-Conference on Human-Level Artificial Intelligence {(HLAI} 2016), New York City, NY, USA, July 16-17, 2016}, volume 1895 of \emph{{CEUR} Workshop Proceedings}, 117--123. CEUR-WS.org.

\bibitem[{Beckers, Chockler, and Halpern(2022)}]{beckers2022causal}
Beckers, S.; Chockler, H.; and Halpern, J. 2022.
\newblock A causal analysis of harm.
\newblock \emph{Advances in Neural Information Processing Systems}, 35: 2365--2376.

\bibitem[{Beckers, Chockler, and Halpern(2023)}]{beckers2023quantifying}
Beckers, S.; Chockler, H.; and Halpern, J.~Y. 2023.
\newblock Quantifying harm.
\newblock In \emph{Proceedings of the Thirty-Second International Joint Conference on Artificial Intelligence}, IJCAI '23.
\newblock ISBN 978-1-956792-03-4.

\bibitem[{Benlamine et~al.(2015)Benlamine, Chaouachi, Villata, Cabrio, Frasson, and Gandon}]{BenlamineCVCFG15}
Benlamine, M.~S.; Chaouachi, M.; Villata, S.; Cabrio, E.; Frasson, C.; and Gandon, F. 2015.
\newblock Emotions in Argumentation: an Empirical Evaluation.
\newblock In Yang, Q.; and Wooldridge, M.~J., eds., \emph{Proceedings of the Twenty-Fourth International Joint Conference on Artificial Intelligence, {IJCAI} 2015, Buenos Aires, Argentina, July 25-31, 2015}, 156--163. {AAAI} Press.

\bibitem[{Berendt et~al.(2021)Berendt, Gandon, Halford, Hall, Hendler, Kinder-Kurlanda, Ntoutsi, and Staab}]{berendt:hal-03189474}
Berendt, B.; Gandon, F.; Halford, S.; Hall, W.; Hendler, J.; Kinder-Kurlanda, K.~E.; Ntoutsi, E.; and Staab, S. 2021.
\newblock {Web Futures: Inclusive, Intelligent, Sustainable The 2020 Manifesto for Web Science}.
\newblock \emph{{Dagstuhl Manifestos}}.

\bibitem[{Board(2022)}]{edpb_guidelines_2022}
Board, E. D.~P. 2022.
\newblock Guidelines 03/2022 on deceptive design patterns in social media platform interfaces: how to recognise and avoid them - v2/0.
\newblock Technical report.

\bibitem[{Bode and Vraga(2015)}]{bode2015erratum}
Bode, L.; and Vraga, E.~K. 2015.
\newblock That Was Wrong: The Correction of MisinformationThrough Related Stories Functionality in Social Media.
\newblock \emph{Journal of Communication}, 65(6): 619–638.

\bibitem[{Bohler(2019)}]{bohler2019bug}
Bohler, S. 2019.
\newblock \emph{Le Bug humain: Pourquoi notre cerveau nous pousse {\`a} d{\'e}truire la plan{\`e}te et comment l'en emp{\^e}cher}.
\newblock Robert Laffont.

\bibitem[{Broadbent et~al.(2024)Broadbent, Forestier, Khamassi, and Zolynski}]{broadbent:hal-04542396}
Broadbent, S.; Forestier, F.; Khamassi, M.; and Zolynski, C. 2024.
\newblock \emph{{Pour une nouvelle culture de l'attention}}.
\newblock {Odile Jacob}.

\bibitem[{Bronner(2021)}]{bronner2021}
Bronner, G. 2021.
\newblock \emph{Apocalypse cognitive}.
\newblock Presses Universitaires de France.
\newblock ISBN 978-2-13-073304-1.

\bibitem[{Carroll and correspondent(2023)}]{carroll_much_2023}
Carroll, R.; and correspondent, R. C.~I. 2023.
\newblock ‘{Much} easier to say no’: {Irish} town unites in smartphone ban for young children.
\newblock \emph{The Guardian}.

\bibitem[{Chuai and Zhao(2022)}]{chuai_anger_2022}
Chuai, Y.; and Zhao, J. 2022.
\newblock Anger can make fake news viral online.
\newblock \emph{Frontiers in Physics}, 10: 970174.

\bibitem[{Cialdini(2008)}]{Cialdini_2008}
Cialdini, R.~B. 2008.
\newblock \emph{Influence: Science and practice}.
\newblock Pearson Education.

\bibitem[{Corbu et~al.(2021)Corbu, Bârgăoanu, Durach, and Udrea}]{corbu_fake_2021}
Corbu, N.; Bârgăoanu, A.; Durach, F.; and Udrea, G. 2021.
\newblock Fake {News} {Going} {Viral}: {The} {Mediating} {Effect} {Of} {Negative} {Emotions}.
\newblock \emph{Media Literacy and Academic Research}, 4(2): 58--87.

\bibitem[{Courbet et~al.(2014)Courbet, Fourquet-Courbet, Kazan, and Intartaglia}]{jcc4.12035}
Courbet, D.; Fourquet-Courbet, M.-P.; Kazan, R.; and Intartaglia, J. 2014.
\newblock The Long-Term Effects of E-Advertising: The Influence of Internet Pop-ups Viewed at a Low Level of Attention in Implicit Memory.
\newblock \emph{Journal of Computer-Mediated Communication}, 19(2): 274--293.

\bibitem[{Coyne et~al.(2017)Coyne, Radesky, Collier, Gentile, Linder, Nathanson, Rasmussen, Reich, and Rogers}]{coyne2017}
Coyne, S.~M.; Radesky, J.; Collier, K.~M.; Gentile, D.~A.; Linder, J.~R.; Nathanson, A.~I.; Rasmussen, E.~E.; Reich, S.~M.; and Rogers, J. 2017.
\newblock Parenting and Digital Media.
\newblock \emph{Pediatrics}, 140(Supplement 2): S112--S116.

\bibitem[{Crockett(2017)}]{crockett_moral_2017}
Crockett, M.~J. 2017.
\newblock Moral outrage in the digital age.
\newblock \emph{Nature Human Behaviour}, 1(11): 769--771.

\bibitem[{DeVito(2017)}]{devito2017editors}
DeVito, M.~A. 2017.
\newblock From editors to algorithms: A values-based approach to understanding story selection in the Facebook news feed.
\newblock \emph{Digital journalism}, 5(6): 753--773.

\bibitem[{Drummond and Sauer(2018)}]{drummond2018video}
Drummond, A.; and Sauer, J.~D. 2018.
\newblock Video game loot boxes are psychologically akin to gambling.
\newblock \emph{Nature human behaviour}, 2(8): 530--532.

\bibitem[{Fabbri et~al.(2022)Fabbri, Wang, Bonchi, Castillo, and Mathioudakis}]{fabbri_rewiring_2022}
Fabbri, F.; Wang, Y.; Bonchi, F.; Castillo, C.; and Mathioudakis, M. 2022.
\newblock Rewiring {What}-to-{Watch}-{Next} {Recommendations} to {Reduce} {Radicalization} {Pathways}.
\newblock In \emph{Proceedings of the {ACM} {Web} {Conference} 2022}, {WWW} '22, 2719--2728. New York, NY, USA: Association for Computing Machinery.
\newblock ISBN 978-1-4503-9096-5.

\bibitem[{Fan et~al.(2014)Fan, Zhao, Chen, and Xu}]{fan_anger_2014}
Fan, R.; Zhao, J.; Chen, Y.; and Xu, K. 2014.
\newblock Anger {Is} {More} {Influential} than {Joy}: {Sentiment} {Correlation} in {Weibo}.
\newblock \emph{PLOS ONE}, 9(10): e110184.

\bibitem[{Fernández, Bellogín, and Cantador(2021)}]{fernandez_analysing_2021}
Fernández, M.; Bellogín, A.; and Cantador, I. 2021.
\newblock Analysing the {Effect} of {Recommendation} {Algorithms} on the {Amplification} of {Misinformation}.
\newblock ArXiv:2103.14748 [cs].

\bibitem[{Fortunato and Pecoraro(2022)}]{fortunato_social_2022}
Fortunato, P.; and Pecoraro, M. 2022.
\newblock Social media, education, and the rise of populist {Euroscepticism}.
\newblock \emph{Humanities and Social Sciences Communications}, 9(1): 1--13.

\bibitem[{Funke and Flamini(2023)}]{funke_guide_nodate}
Funke, D.; and Flamini, D. 2023.
\newblock A guide to anti-misinformation actions around the world.

\bibitem[{Garrett, Nisbet, and Lynch(2013)}]{garrett2013undermining}
Garrett, R.~K.; Nisbet, E.~C.; and Lynch, E.~K. 2013.
\newblock Undermining the corrective effects of media-based political fact checking? The role of contextual cues and na{\"\i}ve theory.
\newblock \emph{Journal of Communication}, 63(4): 617--637.

\bibitem[{Gray et~al.(2023)Gray, Sanchez~Chamorro, Obi, and Duane}]{gray_mapping_2023}
Gray, C.~M.; Sanchez~Chamorro, L.; Obi, I.; and Duane, J.-N. 2023.
\newblock Mapping the {Landscape} of {Dark} {Patterns} {Scholarship}: {A} {Systematic} {Literature} {Review}.
\newblock In \emph{Designing {Interactive} {Systems} {Conference}}, 188--193. Pittsburgh PA USA: ACM.
\newblock ISBN 978-1-4503-9898-5.

\bibitem[{Harris(2019)}]{harris_how_2019}
Harris, T. 2019.
\newblock How {Technology} is {Hijacking} {Your} {Mind} — from a {Former} {Insider}.

\bibitem[{Hefti and Heinke(2015)}]{hefti_economics_2015}
Hefti, A.; and Heinke, S. 2015.
\newblock On the economics of superabundant information and scarce attention.
\newblock \emph{Œconomia. History, Methodology, Philosophy}, (5-1): 37--76.
\newblock Number: 5-1 Publisher: Association Œconomia.

\bibitem[{Hendricks and Vestergaard(2019)}]{hendricks_attention_2019}
Hendricks, V.~F.; and Vestergaard, M. 2019.
\newblock The {Attention} {Economy}.
\newblock In Hendricks, V.~F.; and Vestergaard, M., eds., \emph{Reality {Lost}: {Markets} of {Attention}, {Misinformation} and {Manipulation}}, 1--17. Cham: Springer International Publishing.
\newblock ISBN 978-3-030-00813-0.

\bibitem[{Intelligence(2022)}]{intelligence_social_2022}
Intelligence, G.~T. 2022.
\newblock Social media, algorithms, and populism.

\bibitem[{Jersild(1927)}]{jersild1927mental}
Jersild, A.~T. 1927.
\newblock \emph{Mental set and shift}.
\newblock 89. Columbia university.

\bibitem[{Kantrowitz(2021)}]{Kantrowitz_5_2021}
Kantrowitz, A. 2021.
\newblock 5 {Ways} {China} is {Mandating} {Social} {Media} {Changes}.

\bibitem[{Keib et~al.(2018)Keib, Espina, Lee, Wojdynski, Choi, and Bang}]{keib_picture_2018}
Keib, K.; Espina, C.; Lee, Y.-I.; Wojdynski, B.~W.; Choi, D.; and Bang, H. 2018.
\newblock Picture {This}: {The} {Influence} of {Emotionally} {Valenced} {Images}, {On} {Attention}, {Selection}, and {Sharing} of {Social} {Media} {News}.
\newblock \emph{Media Psychology}, 21(2): 202--221.

\bibitem[{Kitchens, Johnson, and Gray(2020)}]{kitchens_understanding_2020}
Kitchens, B.; Johnson, S.~L.; and Gray, P. 2020.
\newblock Understanding {Echo} {Chambers} and {Filter} {Bubbles}: {The} {Impact} of {Social} {Media} on {Diversification} and {Partisan} {Shifts} in {News} {Consumption}.
\newblock \emph{MIS Quarterly}, 44(4): 1619--1649.

\bibitem[{Kohout, Kruikemeier, and Bakker(2023)}]{kohout_may_2023}
Kohout, S.; Kruikemeier, S.; and Bakker, B.~N. 2023.
\newblock May {I} have your {Attention}, please? {An} eye tracking study on emotional social media comments.
\newblock \emph{Computers in Human Behavior}, 139: 107495.

\bibitem[{Kramer(2012)}]{kramer_spread_2012}
Kramer, A.~D. 2012.
\newblock The spread of emotion via {Facebook}.
\newblock In \emph{Proceedings of the {SIGCHI} {Conference} on {Human} {Factors} in {Computing} {Systems}}, {CHI} '12, 767--770. New York, NY, USA: Association for Computing Machinery.
\newblock ISBN 978-1-4503-1015-4.

\bibitem[{Lewis(2017)}]{lewis_our_2017}
Lewis, P. 2017.
\newblock '{Our} minds can be hijacked': the tech insiders who fear a smartphone dystopia.
\newblock \emph{The Guardian}.

\bibitem[{Martel, Pennycook, and Rand(2020)}]{martel_reliance_2020}
Martel, C.; Pennycook, G.; and Rand, D.~G. 2020.
\newblock Reliance on emotion promotes belief in fake news.
\newblock \emph{Cognitive Research: Principles and Implications}, 5(1): 47.

\bibitem[{Mascarell(2020)}]{mascarell2020bibliometric}
Mascarell, A.~B. 2020.
\newblock A bibliometric analysis of utopian literature.
\newblock \emph{ES Review. Spanish Journal of English Studies}, (41): 77--103.

\bibitem[{Monsell(2003)}]{monsell2003task}
Monsell, S. 2003.
\newblock Task switching.
\newblock \emph{Trends in cognitive sciences}, 7(3): 134--140.

\bibitem[{Mudde and Kaltwasser(2017)}]{mudde2017populism}
Mudde, C.; and Kaltwasser, C.~R. 2017.
\newblock \emph{Populism: A very short introduction}.
\newblock Oxford University Press.

\bibitem[{Newman(2019)}]{newman_regulating_2019}
Newman, J.~M. 2019.
\newblock Regulating {Attention} {Markets}.
\newblock \emph{University of Miami Legal Studies Research Paper}.

\bibitem[{Ocampo et~al.(2023)Ocampo, Sviridova, Cabrio, and Villata}]{ocampo-etal-2023-depth}
Ocampo, N.; Sviridova, E.; Cabrio, E.; and Villata, S. 2023.
\newblock An In-depth Analysis of Implicit and Subtle Hate Speech Messages.
\newblock In \emph{Proceedings of the 17th Conference of the European Chapter of the Association for Computational Linguistics}, 1997--2013. Dubrovnik, Croatia: Association for Computational Linguistics.

\bibitem[{O'neil(2017)}]{oneil2017weapons}
O'neil, C. 2017.
\newblock \emph{Weapons of math destruction: How big data increases inequality and threatens democracy}.
\newblock Crown.

\bibitem[{Pariser(2011)}]{pariser2011filter}
Pariser, E. 2011.
\newblock \emph{The filter bubble: How the new personalized web is changing what we read and how we think}.
\newblock Penguin.

\bibitem[{Parkin(2018)}]{parkin_youtube_2018}
Parkin, S. 2018.
\newblock The {YouTube} stars heading for burnout: ‘{The} most fun job imaginable became deeply bleak’.
\newblock \emph{The Guardian}.

\bibitem[{Patino(2024)}]{patino_lemergence_2024}
Patino, B. 2024.
\newblock L’émergence du caudillo numérique.
\newblock \emph{The Tocqueville Review}.
\newblock Publisher: University of Toronto Press.

\bibitem[{Ra et~al.(2018)Ra, Cho, Stone, De~La~Cerda, Goldenson, Moroney, Tung, Lee, and Leventhal}]{ra_association_2018}
Ra, C.~K.; Cho, J.; Stone, M.~D.; De~La~Cerda, J.; Goldenson, N.~I.; Moroney, E.; Tung, I.; Lee, S.~S.; and Leventhal, A.~M. 2018.
\newblock Association of {Digital} {Media} {Use} {With} {Subsequent} {Symptoms} of {Attention}-{Deficit}/{Hyperactivity} {Disorder} {Among} {Adolescents}.
\newblock \emph{JAMA}, 320(3): 255--263.

\bibitem[{Richards(2007)}]{richards2007emotional}
Richards, B. 2007.
\newblock \emph{Emotional governance: Politics, media and terror}.
\newblock Springer.

\bibitem[{Richards and Gross(2000)}]{richards2000emotion}
Richards, J.~M.; and Gross, J.~J. 2000.
\newblock Emotion regulation and memory: the cognitive costs of keeping one's cool.
\newblock \emph{Journal of personality and social psychology}, 79(3): 410.

\bibitem[{Robertson et~al.(2023)Robertson, Pröllochs, Schwarzenegger, Pärnamets, Van~Bavel, and Feuerriegel}]{robertson_negativity_2023}
Robertson, C.~E.; Pröllochs, N.; Schwarzenegger, K.; Pärnamets, P.; Van~Bavel, J.~J.; and Feuerriegel, S. 2023.
\newblock Negativity drives online news consumption.
\newblock \emph{Nature Human Behaviour}, 7(5): 812--822.

\bibitem[{Rocha et~al.(2023)Rocha, de~Moura, Desidério, de~Oliveira, Lourenço, and de~Figueiredo~Nicolete}]{rocha_impact_2023}
Rocha, Y.~M.; de~Moura, G.~A.; Desidério, G.~A.; de~Oliveira, C.~H.; Lourenço, F.~D.; and de~Figueiredo~Nicolete, L.~D. 2023.
\newblock The impact of fake news on social media and its influence on health during the {COVID}-19 pandemic: a systematic review.
\newblock \emph{Journal of Public Health}, 31(7): 1007--1016.

\bibitem[{Rogers and Monsell(1995)}]{rogers1995costs}
Rogers, R.~D.; and Monsell, S. 1995.
\newblock Costs of a predictible switch between simple cognitive tasks.
\newblock \emph{Journal of experimental psychology: General}, 124(2): 207.

\bibitem[{Rosa and Chaumont(2017)}]{rosa2017alienation}
Rosa, H.; and Chaumont, T. 2017.
\newblock \emph{Ali{\'e}nation et acc{\'e}l{\'e}ration: vers une th{\'e}orie critique de la modernit{\'e} tardive}.
\newblock La d{\'e}couverte.

\bibitem[{Rouvroy and Berns(2013)}]{rouvroy2013gouvernementalite}
Rouvroy, A.; and Berns, T. 2013.
\newblock Gouvernementalit{\'e} algorithmique et perspectives d'{\'e}mancipation.
\newblock \emph{R{\'e}seaux}, 177(1): 163--196.

\bibitem[{Santos, Morozovaite, and De~Conca(2024)}]{santos_no_2024}
Santos, C.; Morozovaite, V.; and De~Conca, S. 2024.
\newblock No harm no foul: how harms caused by dark patterns are conceptualised and tackled under {EU} data protection, consumer and competition laws.

\bibitem[{Sasahara et~al.(2021)Sasahara, Chen, Peng, Ciampaglia, Flammini, and Menczer}]{sasahara_social_2021}
Sasahara, K.; Chen, W.; Peng, H.; Ciampaglia, G.~L.; Flammini, A.; and Menczer, F. 2021.
\newblock Social {Influence} and {Unfollowing} {Accelerate} the {Emergence} of {Echo} {Chambers}.
\newblock \emph{Journal of Computational Social Science}, 4(1): 381--402.
\newblock ArXiv:1905.03919 [physics].

\bibitem[{Schonfeld et~al.(2023)Schonfeld, McNiel, Toyoshima, and Binder}]{schonfeld_cyberbullying_2023}
Schonfeld, A.; McNiel, D.; Toyoshima, T.; and Binder, R. 2023.
\newblock Cyberbullying and {Adolescent} {Suicide}.
\newblock \emph{Journal of the American Academy of Psychiatry and the Law Online}.
\newblock Publisher: Journal of the American Academy of Psychiatry and the Law Online Section: Analysis and Commentary.

\bibitem[{Shah(2022)}]{shah_making_2022}
Shah, N. 2022.
\newblock Making {TikTok} {Videos} {Leaves} {Musicians} {Feeling} {Burnout}.
\newblock \emph{Wall Street Journal}.

\bibitem[{Sharma et~al.(2019)Sharma, Qian, Jiang, Ruchansky, Zhang, and Liu}]{sharma_combating_2019}
Sharma, K.; Qian, F.; Jiang, H.; Ruchansky, N.; Zhang, M.; and Liu, Y. 2019.
\newblock Combating {Fake} {News}: {A} {Survey} on {Identification} and {Mitigation} {Techniques}.
\newblock ArXiv:1901.06437 [cs, stat].

\bibitem[{Siegrist and Cvetkovich(2001)}]{siegrist_better_2001}
Siegrist, M.; and Cvetkovich, G. 2001.
\newblock Better negative than positive? {Evidence} of a bias for negative information about possible health dangers.
\newblock \emph{Risk Analysis: An Official Publication of the Society for Risk Analysis}, 21(1): 199--206.

\bibitem[{Small et~al.(2021)Small, Jorkegren, Erkkilä, Shaw, and MEGILL}]{small_polis_2021}
Small, C.; Jorkegren, M.; Erkkilä, T.; Shaw, L.; and MEGILL, C. 2021.
\newblock Polis: Scaling Deliberation by Mapping High Dimensional Opinion Spaces.
\newblock \emph{RECERCA. Revista de Pensament i Anàlisi}.

\bibitem[{Smith(2018)}]{smith_attention_2018}
Smith, D.~R. 2018.
\newblock Attention, attention: your most valuable scientific assets are under attack.
\newblock \emph{EMBO reports}, 19(3): e45684.
\newblock Publisher: John Wiley \& Sons, Ltd.

\bibitem[{Soroka, Fournier, and Nir(2019)}]{soroka_cross-national_2019}
Soroka, S.; Fournier, P.; and Nir, L. 2019.
\newblock Cross-national evidence of a negativity bias in psychophysiological reactions to news.
\newblock \emph{Proceedings of the National Academy of Sciences}, 116(38): 18888--18892.

\bibitem[{Stark et~al.(2020)Stark, Stegmann, Magin, and J{\"u}rgens}]{stark2020algorithms}
Stark, B.; Stegmann, D.; Magin, M.; and J{\"u}rgens, P. 2020.
\newblock Are algorithms a threat to democracy? The rise of intermediaries: A challenge for public discourse.
\newblock \emph{Algorithm Watch}, 26.

\bibitem[{Suler(2004)}]{suler_online_2004}
Suler, J. 2004.
\newblock The {Online} {Disinhibition} {Effect}.
\newblock \emph{CyberPsychology \& Behavior}, 7: 321--326.

\bibitem[{Talwar et~al.(2019)Talwar, Dhir, Kaur, Zafar, and Alrasheedy}]{talwar_why_2019}
Talwar, S.; Dhir, A.; Kaur, P.; Zafar, N.; and Alrasheedy, M. 2019.
\newblock Why do people share fake news? {Associations} between the dark side of social media use and fake news sharing behavior.
\newblock \emph{Journal of Retailing and Consumer Services}, 51: 72--82.

\bibitem[{{The European Parliament}(2023)}]{european_parliament_addictive}
{The European Parliament}. 2023.
\newblock European {Parliament} resolution of 12 {December} 2023 on addictive design of online services and consumer protection in the {EU} single market (2023/2043({INI})).

\bibitem[{Times(2017)}]{the_economic_times_no_2017}
Times, T.~E. 2017.
\newblock No time to kill: {How} smartphone is pushing chewing gum out of fashion.

\bibitem[{Union(2022)}]{european_union_regulation_nodate}
Union, E. 2022.
\newblock Regulation ({EU}) 2022/2065 of the {European} {Parliament} and of the {Council} of 19 {October} 2022 on a {Single} {Market} {For} {Digital} {Services} and amending {Directive} 2000/31/{EC} ({Digital} {Services} {Act}).
\newblock \emph{Official Journal of the European Union}.

\bibitem[{Venturini(2019)}]{venturini2019fake}
Venturini, T. 2019.
\newblock From fake to junk news: The data politics of online virality.
\newblock In \emph{Data politics}, 123--144. Routledge.

\bibitem[{Villata et~al.(2018)Villata, Benlamine, Cabrio, Frasson, and Gandon}]{VillataBCFG18}
Villata, S.; Benlamine, M.~S.; Cabrio, E.; Frasson, C.; and Gandon, F. 2018.
\newblock Assessing Persuasion in Argumentation through Emotions and Mental States.
\newblock In Brawner, K.; and Rus, V., eds., \emph{Proceedings of the Thirty-First International Florida Artificial Intelligence Research Society Conference, {FLAIRS} 2018, Melbourne, Florida, {USA.} May 21-23 2018}, 134--139. {AAAI} Press.

\bibitem[{Vosoughi, Roy, and Aral(2018)}]{vosoughi_spread_2018}
Vosoughi, S.; Roy, D.; and Aral, S. 2018.
\newblock The spread of true and false news online.
\newblock \emph{Science}, 359(6380): 1146--1151.

\bibitem[{Vraga and Bode(2020)}]{vraga2020correction}
Vraga, E.~K.; and Bode, L. 2020.
\newblock Correction as a solution for health misinformation on social media.
\newblock \emph{American Journal of Public Health}, 110(S3): S278--S280.

\bibitem[{Ward et~al.(2017)Ward, Duke, Gneezy, and Bos}]{ward2017brain}
Ward, A.~F.; Duke, K.; Gneezy, A.; and Bos, M.~W. 2017.
\newblock Brain drain: The mere presence of one’s own smartphone reduces available cognitive capacity.
\newblock \emph{Journal of the Association for Consumer Research}, 2(2): 140--154.

\bibitem[{Waszak, Kasprzycka-Waszak, and Kubanek(2018)}]{waszak_spread_2018}
Waszak, P.~M.; Kasprzycka-Waszak, W.; and Kubanek, A. 2018.
\newblock The spread of medical fake news in social media – {The} pilot quantitative study.
\newblock \emph{Health Policy and Technology}, 7(2): 115--118.

\bibitem[{Whateley(2023)}]{whateley_tiktoks_2023}
Whateley, D. 2023.
\newblock {TikTok}'s music influence is 'exhausting' artists and marketers alike as the industry grapples with the pressure to go viral.

\bibitem[{Whittaker et~al.(2021)Whittaker, Looney, Reed, and Votta}]{whittaker_recommender_2021}
Whittaker, J.; Looney, S.; Reed, A.; and Votta, F. 2021.
\newblock Recommender systems and the amplification of extremist content.
\newblock \emph{Internet Policy Review}, 10(2).

\bibitem[{Zaeem, Li, and Barber(2020)}]{zaeem_sentiment_2020}
Zaeem, R.~N.; Li, C.; and Barber, K.~S. 2020.
\newblock On {Sentiment} of {Online} {Fake} {News}.
\newblock In \emph{2020 {IEEE}/{ACM} {International} {Conference} on {Advances} in {Social} {Networks} {Analysis} and {Mining} ({ASONAM})}, 760--767.
\newblock ISSN: 2473-991X.

\bibitem[{Zendle and Cairns(2018)}]{zendle2018video}
Zendle, D.; and Cairns, P. 2018.
\newblock Video game loot boxes are linked to problem gambling: Results of a large-scale survey.
\newblock \emph{PloS one}, 13(11): e0206767.

\bibitem[{Zolynski, Le~Roy, and Levin(2019)}]{zolynski_leconomie_2019}
Zolynski, C.; Le~Roy, M.; and Levin, F. 2019.
\newblock L'économie de l'attention saisie par le droit.
\newblock \emph{Dalloz IP/IT : droit de la propriété intellectuelle et du numérique}, (11): 614.

\bibitem[{Zomorodi(2017)}]{zomorodi2017bored}
Zomorodi, M. 2017.
\newblock \emph{Bored and brilliant: How time spent doing nothing changes everything}.
\newblock Pan Macmillan.

\bibitem[{Zuboff(2019)}]{zuboff2019age}
Zuboff, S. 2019.
\newblock \emph{The Age of Surveillance Capitalism: The Fight for a Human Future at the New Frontier of Power}.
\newblock Profile Books.
\newblock ISBN 9781781256855.

\end{thebibliography}

\end{document}